\documentclass[aps,prx,onecolumn,superscriptaddress, showpacs]{revtex4-2}

\usepackage[
  left=3.5cm,
  right=3.5cm,
  top=2cm,
  bottom=2cm
]{geometry}

\usepackage{graphicx}
\usepackage{dcolumn}
\usepackage{bm}
\usepackage{color}
\usepackage{braket}
\usepackage{ulem}
\usepackage[usenames,dvipsnames,svgnames,table]{xcolor}
\definecolor{mypurple}{HTML}{2e3092}
\definecolor{myorange}{HTML}{ff7019}
\definecolor{mygreen}{HTML}{00b708}

\usepackage{amsfonts}       % adds mathematical fonts
\usepackage{amsmath}        % adds mathematical environments
\usepackage{bm}             % adds Greek characters in bold
\usepackage{mathrsfs}       % adds a better calligraphic font for some symbols
\usepackage{setspace}       % manages line spacing
\usepackage{enumerate}      % http://ctan.org/pkg/enumerate
\usepackage{soul}           % for highlighting and crossing
\usepackage{xspace}
\usepackage[unicode=true,
            linktocpage,
            linkbordercolor={0.5 0.5 1}, %blue
            citebordercolor={0.5 1 0.5}, %green
            colorlinks = true,
            linkcolor = mypurple,
            citecolor = mygreen,
            urlcolor = mypurple
            ]{hyperref}
\usepackage{changes}
\usepackage{makecell}
\usepackage{lineno}   %This is to add line numbers

\begin{document}
\setcitestyle{super}
% \linenumbers
% Use the \preprint command to place your local institutional report
% number in the upper righthand corner of the title page in preprint mode.
% Multiple \preprint commands are allowed.
% Use the 'preprintnumbers' class option to override journal defaults
% to display numbers if necessary
%\preprint{}

%-----------------------------------------%
%############### Title ###################%
%-----------------------------------------%

\title{Quantum critical origin of strange-metals\\ at the end of a pseudogap phase in infinite-layer nickelates}

%-----------------------------------------%
%############### Authors #################%
%-----------------------------------------%

% repeat the \author .. \affiliation  etc. as needed
% \email, \thanks, \homepage, \altaffiliation all apply to the current
% author. Explanatory text should go in the []'s, actual e-mail
% address or url should go in the {}'s for \email and \homepage.
% Please use the appropriate macro foreach each type of information

% \affiliation command applies to all authors since the last
% \affiliation command. The \affiliation command should follow the
% other information
% \affiliation can be followed by \email, \homepage, \thanks as well.

% \affiliation{Universit\'{e} de Sherbrooke--CNRS and IRL Frontières Quantiques,
% Sherbrooke, Qu\'{e}bec, Canada}

\author{C. Iorio-Duval}
\affiliation{Laboratoire des Solides Irradiés, CEA/DRF/IRAMIS, CNRS, École Polytechnique,
Institut Polytechnique de Paris, F-91128 Palaiseau, France}
\affiliation{Institut quantique, D\'{e}partement de physique  \&  RQMP, Universit\'{e} de Sherbrooke, Sherbrooke,  Qu\'{e}bec, Canada}

\author{E. Beauchesne-Blanchet}
\affiliation{Laboratoire des Solides Irradiés, CEA/DRF/IRAMIS, CNRS, École Polytechnique,
Institut Polytechnique de Paris, F-91128 Palaiseau, France}

\author{F. Perreault}
\affiliation{Laboratoire des Solides Irradiés, CEA/DRF/IRAMIS, CNRS, École Polytechnique,
Institut Polytechnique de Paris, F-91128 Palaiseau, France}

\author{J. L. Santana González}
\affiliation{Laboratoire des Solides Irradiés, CEA/DRF/IRAMIS, CNRS, École Polytechnique,
Institut Polytechnique de Paris, F-91128 Palaiseau, France}

\author{S. Üstün Kaykusuz}
\affiliation{Laboratoire des Solides Irradiés, CEA/DRF/IRAMIS, CNRS, École Polytechnique,
Institut Polytechnique de Paris, F-91128 Palaiseau, France}

\author{W. Sun}
\affiliation{National Laboratory of Solid State Microstructures, Jiangsu Key laboratory of Artificial Functional Materials, College of Engineering and Applied Sciences, Nanjing University,  Nanjing 210093, China}
\affiliation{Collaborative Innovation Center of Advanced Microstructures, Nanjing University, Nanjing 210093, China}

\author{D. Graf}
\affiliation{National High Magnetic Field Laboratory, Tallahassee, FL, USA}

\author{Y. F. Nie}
\affiliation{National Laboratory of Solid State Microstructures, Jiangsu Key laboratory of Artificial Functional Materials, College of Engineering and Applied Sciences, Nanjing University,  Nanjing 210093, China}
\affiliation{Collaborative Innovation Center of Advanced Microstructures, Nanjing University, Nanjing 210093, China}
\affiliation{Jiangsu Physical Science Research center, Nanjing 210093, China}

\author{A. Gourgout}
\affiliation{Laboratoire des Solides Irradiés, CEA/DRF/IRAMIS, CNRS, École Polytechnique,
Institut Polytechnique de Paris, F-91128 Palaiseau, France}

\author{G. Grissonnanche}
\email{gael.grissonnanche@polytechnique.edu}
\affiliation{Laboratoire des Solides Irradiés, CEA/DRF/IRAMIS, CNRS, École Polytechnique,
Institut Polytechnique de Paris, F-91128 Palaiseau, France}

%Collaboration name if desired (requires use of superscriptaddress
%option in \documentclass). \noaffiliation is required (may also be
%used with the \author command).
%\collaboration can be followed by \email, \homepage, \thanks as well.
%\collaboration{}
%\noaffiliation

\date{\today}

% %-----------------------------------------%
% %############### Abstract ################%
% %-----------------------------------------%

\begin{abstract}

\textbf{The quantum-critical origin of strange metals remains debated, particularly in cuprates where $T$-linear resistivity emerges at the end of the pseudogap phase, a regime without long-range order whose nature remains one of the largest mysteries of quantum materials~\cite{Michon2019Thermodynamic, zhong_2022}. Superconducting infinite-layer nickelates provide a new platform to revisit this issue, given their close similarities to cuprates. Here too, $T$-linear resistivity onsets at a critical doping $x^\star$ near the middle of the superconducting dome. Establishing whether $x^\star$ is a quantum critical point (QCP)---a zero-temperature phase transition---typically relies on the electronic specific heat $C_{\rm el}$, which follows $C_{\rm el}/T \propto \log(T)$ at a QCP, rather than the constant behaviour of a conventional metal. However, the thin-film form of infinite-layer nickelates precludes calorimetry. We therefore use the Seebeck coefficient as a low-temperature proxy for specific heat per carrier. In La$_{\rm 1-x}$Sr$_{\rm x}$NiO$_2$ at $x^\star$, the high-temperature Seebeck response is quantitatively captured by the ARPES-measured band structure, indicating well-defined quasiparticles. Below 60 kelvin, however, $S/T$ develops a logarithmic divergence, $S/T \propto \log(T)$, persisting to the lowest temperature once superconductivity is suppressed by $B=41.5$~T. This identifies $x^\star$ as a QCP terminating the underdoped phase. Finally, we find that the carrier density $n_{\rm d}$ of the Ni-$d_{\rm x^2-y^2}$ pocket drops from $1+x$ above $x^\star$ to $x$ below, mirroring the hallmark of the pseudogap phase in cuprates and iridates. These results point to a pseudogap-like underdoped regime ending at $x^\star$, from which strange-metal behaviour emerges.}

\end{abstract}

%-----------------------------------------%
%############### PACS ####################%
%-----------------------------------------%

% % insert suggested PACS numbers in braces on next line
% \pacs{74.72.Gh, 74.25.Dw, 74.25.F-}

% 74.72.Gh  Hole-doped cuprate superconductors
% 74.25.Dw  Phase diagrams superconductivity
% 74.25.F-  Transport properties

%-----------------------------------------%
%############### Keywords ################%
%-----------------------------------------%

% insert suggested keywords - APS authors don't need to do this
%\keywords{}

%-----------------------------------------%
%############### Make Title ##############%
%-----------------------------------------%

%\maketitle must follow title, authors, abstract, \pacs, and \keywords
\maketitle

%-----------------------------------------%
%############### MAIN ####################%
%-----------------------------------------%

% body of paper here - Use proper section commands
% References should be done using the \cite,We synthesized t, \ref, and \label commands

%>>>>>>>>>>>>>>>>>>>>>>>>>>>>>>>>>>>>>>>>>>>>>>>>>>>>>>>>>>>>>>>>>>>>>>>>>>>>>>>>>>>>>>>>>>>>>>>>>>
\section{INTRODUCTION}
The recurring pattern of strange metals in quantum materials, with persisting $T$-linear resistivity defies the conventional theory of metals and remains one of the most challenging problem of condensed matter research~\cite{hartnoll_2022,patel_2019}. In addition, strange metals tend to precede high-temperature superconductivity, indicating a common origin between the strange scattering of the electrons in the normal state and their pairing, reinforcing the urge to an understanding~\cite{taillefer_2010,yuan_2022}.

Often, if not always, strange metals seem to appear at a quantum critical point (QCP) -- a zero-temperature phase transition tuned by non-thermal parameters such as doping -- that marks the end of a phase. It has been largely documented in heavy-fermion~\cite{bianchi_2003}, iron-based~\cite{walmsley_2013} and organic~\cite{wakamatsu_2023,doiron-leyraud_2009} superconductors, as well as ruthenates~\cite{rost_2011}, and magic angle twisted bilayer graphene~\cite{jaoui_2022}. At a quantum critical point, the electronic specific heat diverges logarithmically in temperature $C_{\rm el}/T\propto \log(T)$~\cite{lohneysen_1994,rost_2011,Michon2019Thermodynamic}, because the entropy keeps rising as the material reaches the $T\rightarrow 0$ limit -- chasing a phase transition that would never happen. The quantum critical material becomes scale invariant, which means that the temperature is, at that point, the only energy scale of the system~\cite{hartnoll_2022,varma_2020b,hu_2024}, leading to a timescale $\tau \sim \hbar / k_{\rm B}T/\hbar$. In this scenario, the scattering rate is naturally bounded at the Planckian limit~\cite{Bruin2013Similarity,Legros2019Universal,Grissonnanche2021LinearIn}, resulting in the characteristic $T$-linear resistivity of strange metals. Establishing experimentally that strange metals systematically occur at quantum critical points would provide compelling evidence for their quantum-critical origin and a natural framework for understanding their anomalous transport properties.

\begin{figure}[t]
\centering
\includegraphics[width=0.9\textwidth]{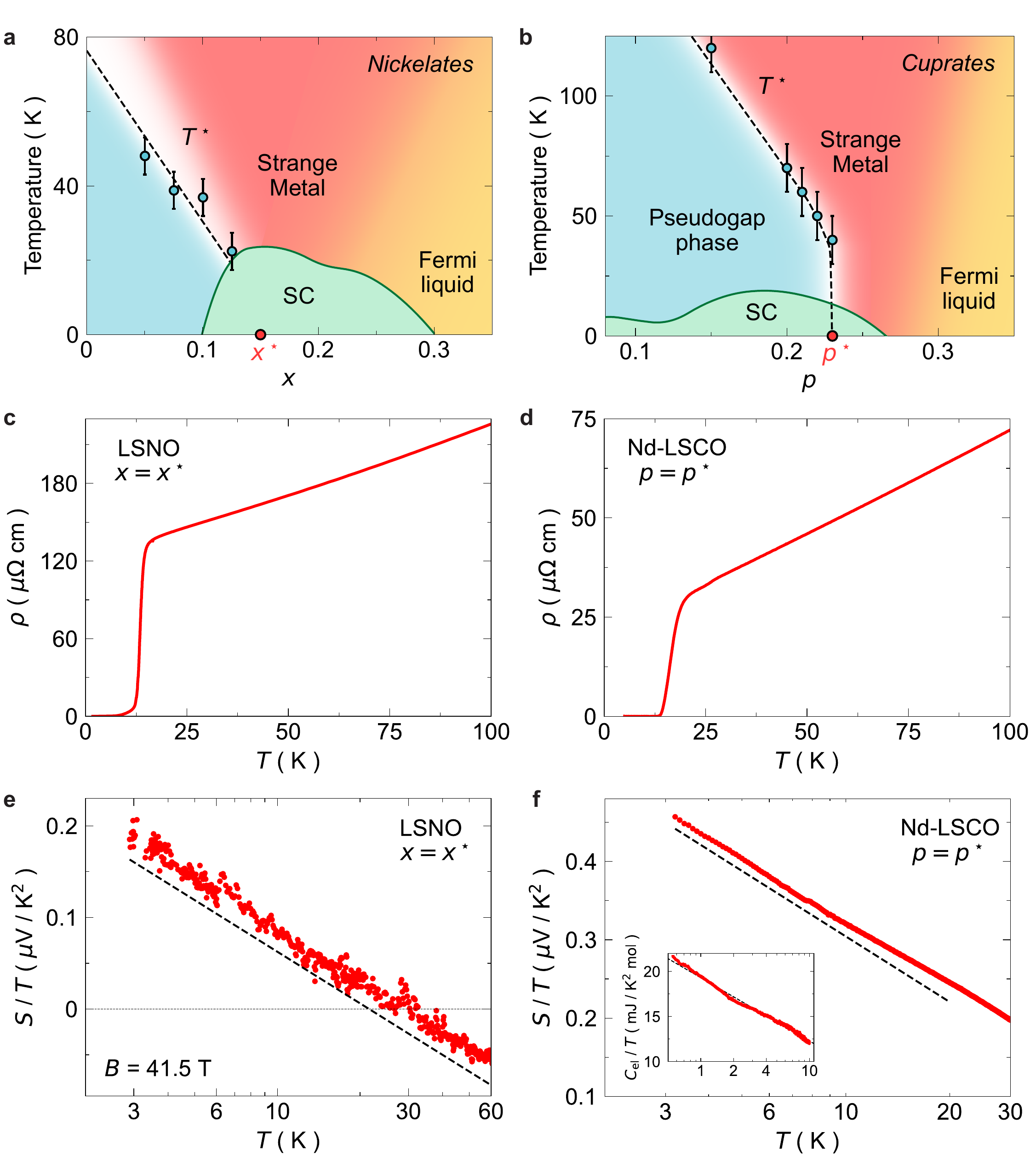}
\caption{
Comparisons between infinite-layer nickelates (left) and hole doped cuprates. (\textbf{a,b}) Temperature $T$ versus hole doping $x$ ($p$) phase diagram -- $x^\star$ ($p^\star$) represents the onset doping of the $T$-linear resistivity: (a) infinite-layer nickelate Nd$_{\rm 1-x}$Sr$_{\rm x}$NiO$_2$ (NSNO) -- reproduced from Lee~\textit{et al.}~\cite{lee_linear_character_2023} and (b) hole-doped cuprate (Nd$_{0.4}$)La$_{\rm 1.6-x}$Sr$_{\rm x}$CuO$_4$ (Nd-LSCO/LSCO), with $p=x$ -- reproduced from Collignon~\textit{et al.}~\cite{collignon_2017}. $T^\star$ (blue points) marks the deviation from a linear in temperature behavior towards an upturn in resistivity. Lines are a guide to the eye; (\textbf{c,d}) In-plane resistivity $\rho$ vs $T$ in the strange metal regime measured at zero field of (c) La$_{\rm 1-x}$Sr$_{\rm x}$NiO$_2$ (LSNO) at $x=x^\star$ and (d) Nd-LSCO at $p=p^\star$ ($p=0.24$) (data reproduced from Daou~\textit{et al.}~\cite{Daou2009Linear}); (\textbf{e,f}) Seebeck coefficient plotted as $S/T$ vs $T$ on logarithmic scale for (e) LSNO at $x=x^\star$, and (f) Nd-LSCO at $p=p^\star$ (data reproduced from Gourgout~\textit{et al.}~\cite{Gourgout2022Seebeck}).  The inset in (f) shows the electronic specific heat for Nd-LSCO at $p=p^\star$ plotted as $C_{\rm el}/T$ vs $T$ on a logarithmic scale (data reproduced from Michon~\textit{et al.}~\cite{Michon2019Thermodynamic}). Dashed lines are guide for the eye.
}
\label{fig:phase_diagram}
\end{figure}

However, the universality of the link between strange metals and quantum criticality has been questioned in hole-doped cuprates, where $T$-linear resistivity emerges at the endpoint $p^\star$ of the pseudogap phase (Fig.~\ref{fig:phase_diagram}d)---a regime without long-range magnetic order whose nature remains unresolved (Fig.~\ref{fig:phase_diagram}b). While $p^\star$ was proposed to be a genuine QCP, based on the logarithmic divergence $C_{\rm el}/T\propto \log(T)$ measured in Eu/Nd$_{0.4}$La$_{\rm 1.6-x}$Sr$_{\rm x}$CuO$_4$ (Eu/Nd-LSCO) (Fig.~\ref{fig:phase_diagram}f)~\cite{Michon2019Thermodynamic}, it was also argued that this divergence could instead reflect the nearby van Hove singularity in this family~\cite{zhong_2022}---a far more conventional band-structure effect than quantum criticality, and one with much less relevance to the broader problem of strange metals and high-temperature superconductivity.

Superconducting infinite-layer nickelates~\cite{Li2019Superconductivity,wang_2024,wang_2025} provide a timely platform to revisit this question, as they reproduce several key ingredients of the cuprates, including nodal superconductivity~\cite{Ranna2025,cheng_2024}, a correlated Ni-$d_{x^2-y^2}$ band at the Fermi level~\cite{Sun2025Electronic}, and a high-doping Fermi-liquid regime~\cite{lee_linear_character_2023}. Crucially, they enter a strange-metal regime with $T$-linear resistivity at an onset doping $x^\star$ located near the middle of the superconducting dome~\cite{lee_linear_character_2023}, reminiscent of the cuprate phenomenology (Fig.~\ref{fig:phase_diagram}a,b). This makes nickelates a new setting in which to test whether strange-metal behavior is tied to quantum criticality. Yet, direct evidence that $x^\star$ corresponds to a quantum critical point is still lacking, and the nature of the low-doping phase that terminates at $x^\star$ remains unresolved (Fig.~\ref{fig:phase_diagram}a).

A way to test whether $T$-linear resistivity in infinite-layer nickelates (Fig.~\ref{fig:phase_diagram}c) emerges from quantum criticality is to track how the entropy evolves with temperature. However, since these materials can only be stabilized as thin films ($<10$ nm)~\cite{suyolcu_2025}, conventional calorimetry is ineffective here to track the evolution of entropy, as it primarily measures the response of the underlying substrate. To circumvent this limitation, we turn to the Seebeck coefficient $S$, which at low temperatures can be regarded as the specific heat per carrier -- with $S \sim C_{\rm el} / n e$, $n$ the carrier density and $e$ the electronic charge. In cuprates, both $C_{\rm el}/T$~\cite{Michon2019Thermodynamic} and $S/T$~\cite{Gourgout2022Seebeck} show the same logarithmic divergence in the strange metal regime and is supported by theory~\cite{varma_2020b,bashan_2026} -- this behavior is also observed in electron-doped cuprates~\cite{mandal_2019}, heavy fermions~\cite{hartmann_2010}, iron-based~\cite{gooch_2009} and organic superconductors~\cite{wakamatsu_2023} -- demonstrating that the Seebeck effect can be used as a proxy for the $log(T)$ dependence of the specific heat at a QCP (Fig.~\ref{fig:phase_diagram}f). In this work, we find that strange metal infinite-layer nickelates at the onset doping $x^\star$ of the strange metal behavior display the same signature: once superconductivity has been suppressed with a large magnetic field, $S/T$ is found to diverge logarithmically $S/T \propto \log(T)$ as $T \rightarrow 0$ in the normal state (Fig.~\ref{fig:phase_diagram}e). This behavior is only found at $x^\star$. Therefore, we conclude that the onset doping of $T$-linear resistivity $x^\star$ is a quantum critical point.

We then turn to the nature of the low-doping phase below $x^\star$ in infinite-layer nickelates and ask whether it is analogous to the pseudogap phase of hole-doped cuprates. In cuprates, the defining phenomenology of the pseudogap phase is a loss of charge carriers below a crossover temperature $T^\star$, revealed by an upturn in the resistivity and by a sudden change in the zero-temperature Hall coefficient $R_{\rm H}(T\rightarrow 0)$ as a function of doping. This change reflects a drop in the carrier density of the Cu-$d_{x^2-y^2}$ states from $1+p$ holes per Cu above the critical doping $p^\star$ to $p$ below $p^\star$~\cite{badoux2016,collignon_2017}. Strikingly, this occurs without the onset of long-range antiferromagnetic order below $p^\star$.

In nickelates, both the resistivity upturn upon cooling and the sudden change in the Hall coefficient below $x^\star$ are observed~\cite{lee_linear_character_2023}, despite the absence of long-range order in the low-doping regime. However, the multiband nature of infinite-layer nickelates complicates the direct observation of the evolution of the carrier-density $n_{\rm d}$ in the Ni-$d_{x^2-y^2}$ band across the QCP. Using a two-band analysis constrained by the angle-resolved photoemission spectroscopy (ARPES) band structure~\cite{Sun2025Electronic}, we find that $n_{\rm d}$ collapses from $1+x$ above $x^\star$ to $x$ below, mirroring the hallmark signature of the pseudogap phase in cuprates. We therefore conclude that the low-doping regime of infinite-layer nickelates is a pseudogap-like state that terminates at the QCP $x^\star$, from which the $T$-linear resistivity onsets, reinforcing the quantum-critical origin of strange metals.

%#################### Figure 1 ####################%
%##################################################%

%>>>>>>>>>>>>>>>>>>>>>>>>>>>>>>>>>>>>>>>>>>>>>>>>>>>>>>>>>>>>>>>>>>>>>>>>>>>>>>>>>>>>>>>>>>>>>>>>>>
\section{RESULTS}

\begin{figure}[t]
\centering
\includegraphics[width=1\textwidth]{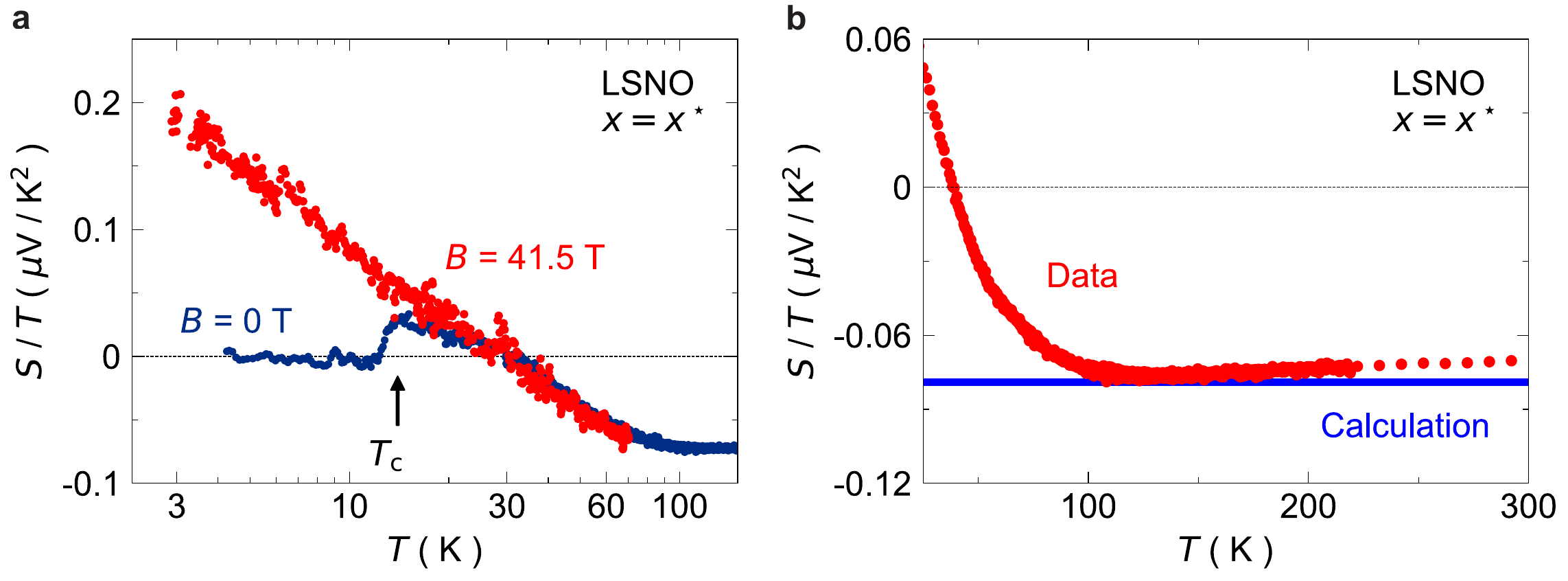}
\caption{
(\textbf{a}) Seebeck coefficient plotted as $S/T$ vs $T$ of LSNO at $x=x^\star$ on a logarithmic scale (sample S1) for $B=0$~T (blue points) and $B=41.5$~T (red points). (\textbf{b}) $S/T$ vs $T$ for LSNO $x=x^\star$: data (sample S2, red points) are compared to Boltzmann transport Seebeck calculations (blue line).
}
\label{fig:seebeck_log_boltz}
\end{figure}

\textbf{Seebeck effect}.
To determine whether the onset of the $T$-linear resistivity in infinite-layer nickelates corresponds to a quantum critical point, we measure the Seebeck coefficient $S$ as a function of temperature in La$_{1-x}$Sr$_x$NiO$_2$ (LSNO) at $x^\star$ (for exact doping value, see Methods) (Fig.~\ref{fig:phase_diagram}c). The data reveal two distinct regimes: above $60$~K, $S/T$ is constant and negative, whereas below this temperature it rises sharply, changes sign, and diverges upon cooling (Fig.~\ref{fig:seebeck_log_boltz}).

We first focus on the temperature regime above $60$~K. In a metal with well-defined quasiparticles, $S/T$ is governed by the particle–hole asymmetry of the electronic band dispersion and is therefore expected to be temperature-independent in the most basic description. A negative, constant $S/T$ over a broad temperature range, with a comparable magnitude, has been reported in related systems such as the five-layer nickelate Nd$_6$Ni$_5$O$_{12}$ and the cuprate (Bi,Pb)$_2$(Sr,La)$_2$CuO$_{6+\delta}$~\cite{grissonnanche2024,Kondo2005Contribution}. In these materials, Boltzmann transport calculations based on ARPES-derived or density-functional-theory (DFT) band structures quantitatively reproduce the measured Seebeck coefficient, demonstrating that $S/T$ directly reflects the electronic dispersion at the Fermi level~\cite{Gourgout2022Seebeck,grissonnanche2024}.

Applying the same approach to LSNO at $x^\star$, we compute $S/T$ using the experimentally measured ARPES electronic dispersion~\cite{Sun2025Electronic} within Boltzmann transport theory, following the method of Grissonnanche \textit{et al.}~\cite{grissonnanche2024}. The calculation reproduces both the sign and magnitude of the measured $S/T$ with quantitative accuracy (Fig.~\ref{fig:seebeck_log_boltz}b), comparable to overdoped cuprates and multilayer nickelates that share a similar $d_{\rm x^2-y^2}$ band structure~\cite{grissonnanche2024}. These results establish that the high-temperature Seebeck coefficient in LSNO is set by the underlying electronic band structure, indicating well-defined quasiparticles above $60$~K despite the presence of $T$-linear resistivity.

%#################### Figure 2 ####################%
%##################################################%

To access the low-temperature behaviour of the Seebeck coefficient in the normal state of LSNO at $x=x^\star$, we suppress superconductivity using a large magnetic field of $B=41.5$~T, sufficient to reach the normal state down to $T\rightarrow 0$. Below $60$~K, $S/T$ exhibits a clear logarithmic divergence, $S/T \propto \log(T)$ (Fig.~\ref{fig:seebeck_log_boltz}a), over more than a decade in temperature -- a hallmark signature of quantum criticality. This establishes that the onset doping $x^\star$ of the strange-metal regime coincides with a quantum critical point beneath the superconducting dome.

\begin{figure}[h]
\centering
\includegraphics[width=1\textwidth]{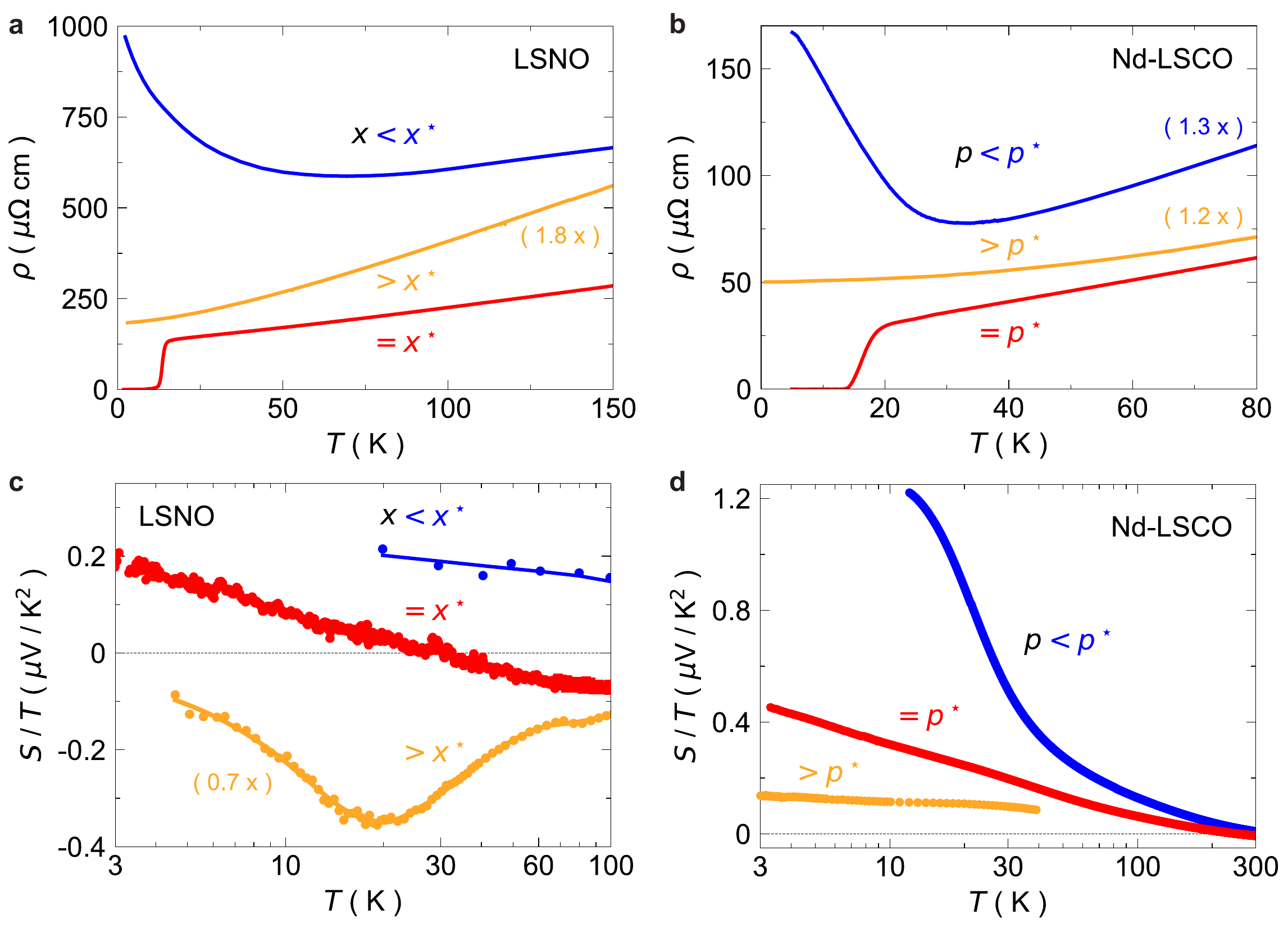}
\caption{
Transport coefficients for doping on each side of the quantum critical point for infinite-layer nickelates (left) and hole doped cuprates (right): underdoped (blue), at the critical doping (red) and overdoped (green). (\textbf{a,b}) In-plane resistivity $\rho$ vs $T$ at $B=0$~T. (a) Infinite-layer nickelate data of LSNO at $x=x^\star$ (sample S2, red line), $x>x^\star$ (0.25, orange line), and $x<x^\star$ (0.04, blue line) (the last doping is reproduced from~\cite{Osada2025Systematic}). (b) Cuprate data of Nd-LSCO at $p<p^\star$ (0.22, blue line), $p=p^\star$ (0.24, red line) and LSCO $p>p^\star$ (0.33, orange line) (reproduced respectively from \cite{collignon_2017,Daou2009Linear,nakamae_2003}). (\textbf{c,d}) Seebeck coefficient plotted as $S/T$ vs $T$ on a logarithmic scale at different doping. (c) Infinite-layer nickelate data of LSNO at $x=x^\star$ (sample S1) at $B=0$~T (red squares) and $B=41.5$~T (red circles), $x>x^\star$ (0.25, orange points) at $B=0$~T, and $x<x^\star$ (0.04, blue points) at $B=0$~T (the last one is reproduced from~\cite{Osada2025Systematic}). (d) Cuprate data of Nd-LSCO at $p<p^\star$ at $B=16$~T (0.22, blue points), $p=p^\star$ at $B=16$~T (0.24, red points) and LSCO $p>p^\star$ at $B=1$~T (0.33, orange line) are reproduced from \cite{Gourgout2022Seebeck,jin2021}. Scaling factors for some curves were used for clarity.
}
\label{fig:comparison_doping}
\end{figure}

To test whether this behaviour is confined to $x^\star$, we examine neighbouring dopings. Measurements at $x>x^\star$ and $x<x^\star$~\cite{Osada2025Systematic}, show no logarithmic divergence of $S/T$ at low temperature (Fig.~\ref{fig:comparison_doping}c). Instead, in overdoped LSNO at $x>x^\star$, $S/T$ saturates at low temperature and recovers at high temperature a value comparable to that observed at $x=x^\star$ in the quasiparticle regime, consistent with conventional metallic behaviour away from criticality. The absence of a $\log(T)$ divergence on both sides of $x^\star$ demonstrates that quantum criticality is confined to a narrow doping range centred at $x^\star$.

%#################### Figure 3 ####################%
%##################################################%

% >>>>>>>>>>>>>>>>>>>>>>>>>>>>>>>>>>>>>>>>>>>>>>>>>>>>>>>>>>>>>>>>>>>>>>>>>>>>>>>>>>>>>>>>>>>>>>>>>
\section{DISCUSSION}

%\onecolumngrid % important to put a space below that statement
%\twocolumngrid

\textbf{Origin of the quantum critical point.}
In infinite-layer nickelates, the QCP at $x^\star$ separates two distinct metallic regimes. On the overdoped side ($x>x^\star$), the in-plane resistivity displays strange-metal behaviour that smoothly evolves into a Fermi-liquid-like regime upon further doping, without a phase transition (Fig.~\ref{fig:comparison_doping}a). Consistently, the Hall coefficient extrapolated to zero temperature is positive, $R_{\rm H}(T\rightarrow 0)>0$~\cite{lee_linear_character_2023} (Fig.~\ref{fig:hall_effect_0_vs_x}). On the underdoped side ($x<x^\star$), the resistivity develops an upturn below a crossover temperature $T^\star$ (Fig.~\ref{fig:comparison_doping}a), while the zero-temperature Hall coefficient becomes negative, $R_{\rm H}(T\rightarrow 0)<0$~\cite{lee_linear_character_2023} (Fig.~\ref{fig:hall_effect_0_vs_x}). Importantly, the resistivity remains finite as $T\rightarrow 0$, indicating that the upturn does not signal an insulating ground state, but rather a transition between two metallic states with different effective carrier densities. In this sense, transport points to a loss of carriers upon reducing doping below $x^\star$ and cooling below $T^\star$, as commonly expected when an ordered phase induces a Fermi-surface reconstruction. Yet, to date, no long-range order has been established in the low-doping regime of infinite-layer nickelates, although signatures of short-range antiferromagnetism have been reported~\cite{lu_magnetic_2021}.

\begin{figure}[h]
\centering
\includegraphics[width=1\textwidth]{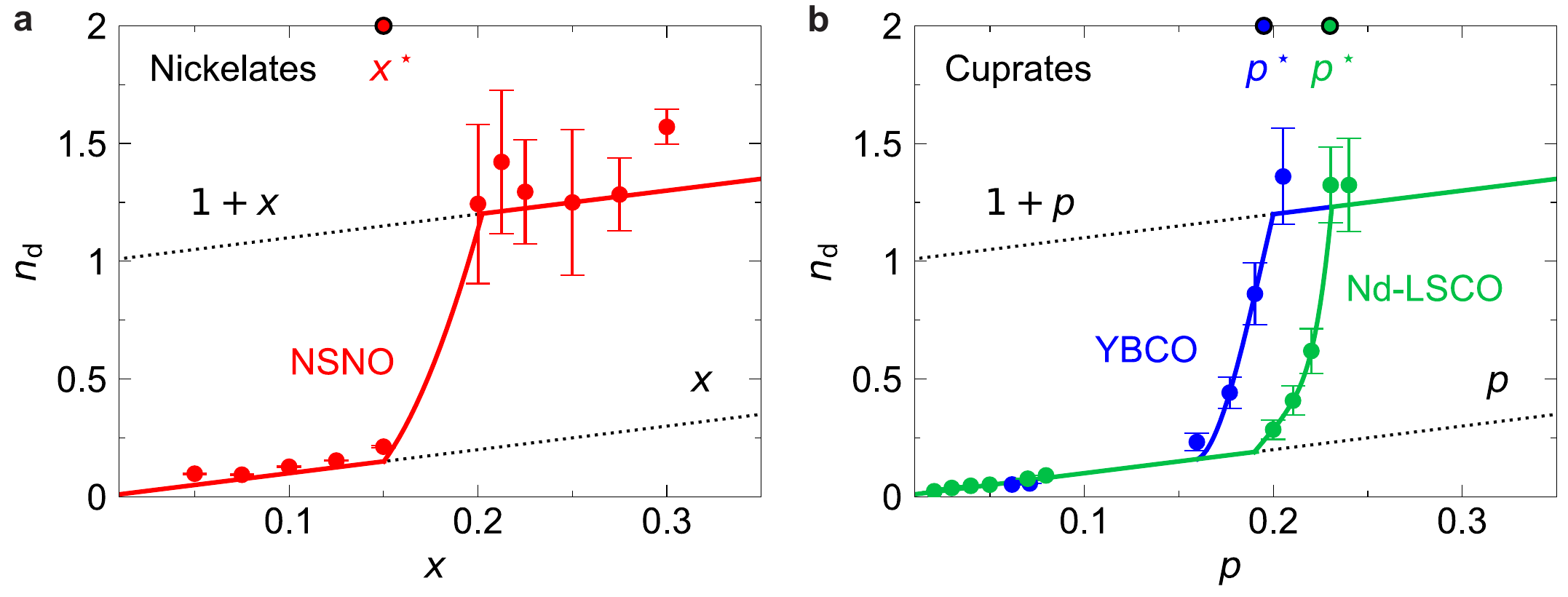}
\caption{
Carrier density as a function of doping across the quantum critical point in infinite-layer nickelates (NSNO, right) and cuprates (YBCO, Nd-LSCO, left) (\textbf{a}) Carrier density $n_{\rm d}$ of the Ni-$d_{\rm x^2-y^2}$ band (red points) as a function of hole doping $x$ extracted from the Hall coefficient $R_{\rm H}(T\rightarrow 0)$ of NSNO~\cite{lee_linear_character_2023} (Fig.~\ref{fig:hall_effect_0_vs_x}) using a two-band model (Equation~\ref{eq:RH_reduced}). The red line is a guide to the eye and the dashed lines represent the $x$ and $1+x$ holes per Ni values; (\textbf{b}) Carrier density $n_{\rm d}$ of the Cu-$d_{\rm x^2-y^2}$ band as a function of doping $p$ extracted from the Hall coefficient using a single band model for YBCO (blue points), Nd-LSCO (green points above $p=0.1$), and LSCO (green points below $p=0.1$) (data reproduced respectively from Badoux~\textit{et al.}~\cite{badoux2016} and Collignon~\textit{et al.}~\cite{collignon_2017}). Solid lines are guides to the eye and the dashed lines represent the $p$ and $1+p$ holes per Cu values.
}
\label{fig:carrier_density_vs_doping}
\end{figure}

An upturn in the resistivity and a rapid change in the Hall coefficient are hallmark signatures of hole-doped cuprates upon entering the pseudogap regime below a crossover temperature $T^\star$ and below a critical doping $p^\star$~\cite{Daou2009Linear} (Fig.~\ref{fig:comparison_doping}a). In these materials, the pseudogap phase is characterized by a loss of itinerant carriers without the onset of long-range order~\cite{Grissonnanche2023No}. Because, only one band crosses the Fermi level in cuprates---the Cu-$d_{x^2-y^2}$ band---the carrier density can be inferred directly from the zero-temperature Hall coefficient via $R_{\rm H}(T\rightarrow 0)=1/(ne)$. Hall data then reveal a sharp reduction in carrier density below $p^\star$~\cite{badoux2016,collignon_2017}, from $n=1+p$ holes per Cu above $p^\star$ to $n=p$ holes per Cu below $p^\star$ (Fig.~\ref{fig:carrier_density_vs_doping}b). This loss of one hole per Cu is what is expected when the system enters an antiferromagnetic state, suggesting a close connection between the pseudogap phase and antiferromagnetism~\cite{Storey2016Hall}.

Assessing whether an analogous carrier-density change occurs in infinite-layer nickelates across $x^\star$ is less straightforward using $R_{\rm H}(T\rightarrow 0)$, because two bands cross the Fermi level: a quasi-two-dimensional, hole-like Ni-$d_{x^2-y^2}$ band with carrier density $n_{\rm d}$, and a three-dimensional, electron-like rare-earth Nd-$s$ band with carrier density $n_{\rm s}$, located respectively near the center and the corners of the first Brillouin zone. Both theory~\cite{karp_2020,chen_2022a} and experiment~\cite{hepting_2020,li_ARPES2025} indicate that the carriers in the 2D $d$ band are strongly correlated, whereas those in the 3D $s$ band are comparatively weakly correlated. We therefore adopt, as a first approximation, that the Nd-$s$ pocket evolves smoothly through $x^\star$ and take $n_{\rm s}$ to be doping independent; we then use this assumption to infer the evolution of $n_{\rm d}$ across the QCP.

%#################### Figure 4 ####################%
%##################################################%

To quantify this carrier loss, we extract the carrier density $n_{\rm d}$ of the Ni-$d_{x^2-y^2}$ pocket from the zero-temperature Hall coefficient $R_{\rm H}(T\rightarrow 0)$ in Nd$_{1-x}$Sr$_x$NiO$_2$ (NSNO), using the Hall data reported by Lee \textit{et al.}~\cite{lee_linear_character_2023} (Fig.~\ref{fig:hall_effect_vs_T}). We employ a two-band model comprising a hole-like $d$ band and an electron-like rare-earth $s$ band. While such a model has been shown to capture the temperature dependence of $R_{\rm H}$ in overdoped infinite-layer nickelates~\cite{Osada2025Systematic}, here we restrict the analysis to the $T\rightarrow 0$ limit (see Methods). We find that $n_{\rm d}$ drops abruptly across the QCP, from $1+x$ above $x^\star$ to $x$ just below $x^\star$ (Fig.~\ref{fig:carrier_density_vs_doping}a). This naturally accounts for the sign change to negative $R_{\rm H}(T\rightarrow 0)$, as the hole-like $d$ contribution becomes subdominant to the electron-like $s$ pocket below $x^\star$. The resulting collapse of $n_{\rm d}$ mirrors the carrier-density drop observed across the pseudogap critical point in cuprates~\cite{badoux2016,collignon_2017} and in iridates~\cite{hsu_2024a}.

This sudden drop in carrier density is puzzling in the absence of established long-range order. Several studies report signatures of short-range magnetism at low doping~\cite{lu_Science_2021}, with phenomenology reminiscent of the cuprate pseudogap phase~\cite{cui_2021, zhao_2021}. In cuprates, the pseudogap is variously discussed in terms of antiferromagnetic correlations~\cite{Storey2016Hall}, topological order~\cite{eberlein_2016}, or a fractionalized Fermi liquid~\cite{sachdev_2025}. Therefore, the resistivity upturn and the drop in the carrier density of Ni-$d_{x^2-y^2}$ carriers below $x^\star$, despite the absence of long-range order, point to a metal-to-metal transition driven by a partial suppression of the Ni-$d_{x^2-y^2}$ degrees of freedom, closely paralleling the defining transport phenomenology of the pseudogap phase in cuprates. We conclude that the underdoped regime of infinite-layer nickelates is a pseudogap-like phase that terminates at the quantum critical point $x^\star$.

The observation of quantum criticality at the endpoint $x^\star$ of a pseudogap-like phase in infinite-layer nickelates reinforces the interpretation of $p^\star$ in cuprates as a genuine quantum critical point and supports a universal connection between quantum criticality and $T$-linear resistivity in both families. In the hole-doped cuprate Nd-LSCO, a logarithmic divergence of both the Seebeck coefficient and the electronic specific heat was reported at $p^\star$~\cite{Gourgout2022Seebeck,Michon2019Thermodynamic}. It was later suggested that this behavior could instead arise from the nearby van Hove singularity, which occurs at a similar doping in that material~\cite{zhong_2022}, although counterarguments have been put forward~\cite{Horio2018ThreeDimensional}. In infinite-layer nickelates, no such van Hove singularity is present at $x^\star$~\cite{Sun2025Electronic}, yet the phenomenology and quantum-critical signatures closely parallel those of cuprates. Taken together, our results strongly support the view that strange-metal physics is driven by quantum-critical fluctuations, leading to $T$-linear resistivity and scattering at the Planckian limit~\cite{varma_2020b,hu_2024}.

\bibliography{references}

%-----------------------------------------%
%############### Appendix ################%
%-----------------------------------------%

% Specify following sections are appendices. Use \appendix* if there
% only one appendix.
\appendix

% %-----------------------------------------%
% %############### Methods ##################%
% %-----------------------------------------%

\section{Methods}
\textbf{Thin film growth.}
La$_{\rm 1-x}$Sr$_{\rm x}$NiO$_2$ (LSNO) perovskite thin films were grown with molecular-beam epitaxy on TiO$_2$-terminated (001)-oriented SrTiO$_3$ substrates. The substrate temperature was kept at $600^{\circ}$C and the background pressure at $10^{-5}$ torr using distilled ozone as the oxidizing agent. LaNiO$_3$ and SrTiO$_3$ were grown using the co-deposition method on SrTiO$_3$  substrates while the precise fluxes were determined based on the reflection high-energy electron diffraction (RHEED) oscillations. After growth, the perovskite films were reduced using CaH$_2$ for 120 min at $400^{\circ}$C, with ramping rate of $10^{\circ}$C/min.

\begin{figure}[h]
\centering
\includegraphics[width=1\textwidth]{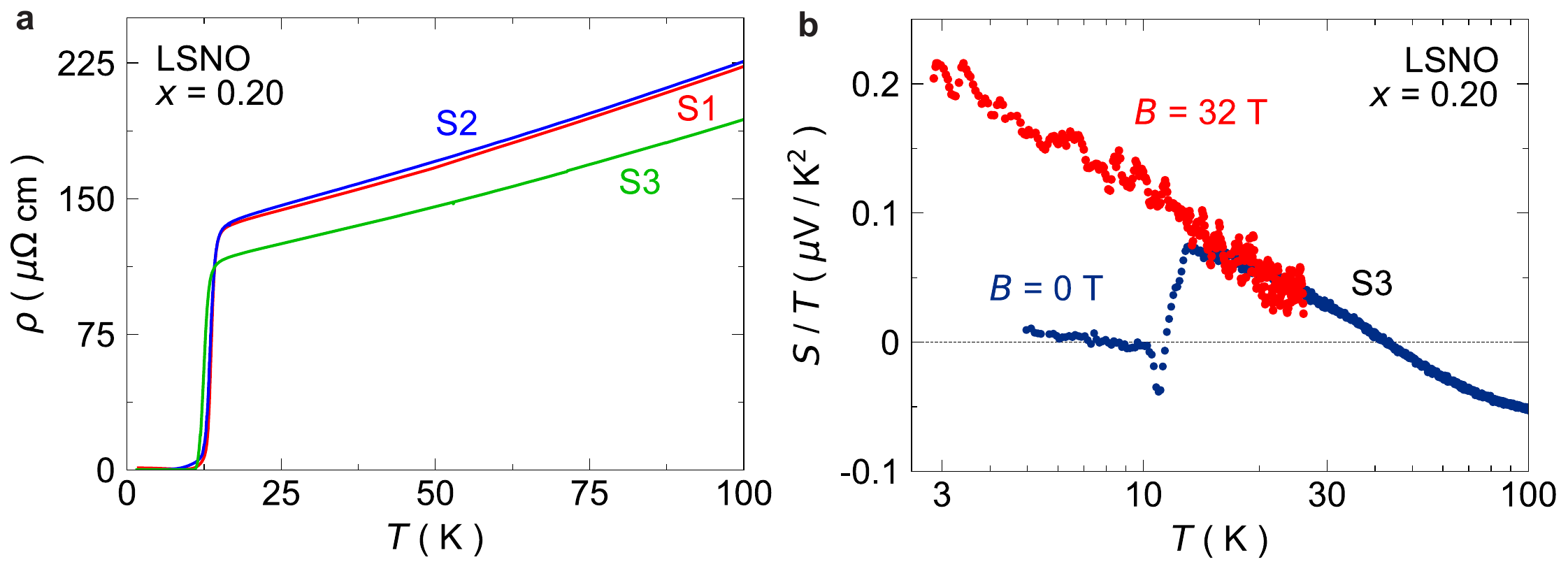}
\caption{
Transport coefficient of different La$_{\rm 1-x}$Sr$_{\rm x}$NiO$_2$ $x = 0.20$ (LSNO) samples used for this study. (\textbf{a}) Comparison of the in-plane resistivity $\rho$ vs $T$ measured at zero field of sample S1 (red), S2 (blue) and S3 (green). (\textbf{b}) Seebeck coefficient plotted as $S/T$ vs $T$ of sample S3 for $B=0$~T (blue points) and $B=32$~T (red points).
}
\label{fig:comparison_samples}
\end{figure}

\textbf{Samples.} Three different LSNO samples at $x=0.20$ were measured in this study. The $x=0.20$ doping marks the onset of $T$-linear resistivity, labeled as $x^\star$ in the rest of the manuscript. All $x=0.20$ samples were prepared the same way and all show similar resistivities (Fig.~\ref{fig:comparison_samples}a). The Seebeck effect was only measured at $B=41.5$~T at the NHMFL for S1 and S3. One LSNO $x=0.25$ sample was also measured and labeled as $x>x^\star$ in Fig.~\ref{fig:comparison_doping}. This sample was prepared using the same methods as the $x=0.20$ sample.

\textbf{Sr concentration and comparison across groups.} We note that the nominal Sr concentration $x$ reported for infinite-layer nickelates can vary systematically between different groups, reflecting intrinsic differences between growth techniques -- in particular molecular beam epitaxy (MBE) and pulsed laser deposition (PLD) -- as well as reduction conditions. Consequently, their seem to be an inherent uncertainty concerning the exact doping level in those thin films.

In the present work, we determine the critical doping $x^\star$ from the onset of $T$-linear resistivity in infinite-layer nickelates. In our MBE-grown LSNO films, we find $x^\star = 0.20$. In contrast, for PLD-grown samples of the same compound, Osada \textit{et al.}~\cite{Osada2025Systematic} report $x^\star \approx 0.14\text{--}0.15$. Similarly, for PLD-grown Nd$_{\rm 1-x}$Sr$_{\rm x}$NiO$_2$, Lee \textit{et al.}~\cite{lee_linear_character_2023} find $x^\star = 0.15$.

\textbf{Contacts.} Electrical contacts were made with ultrasonic wire-bonding and aluminum wires directly on the surface of the film.

\textbf{Electric measurements.}
Standard four-point configuration contacts was used on the samples to measure resistivity. Measurements were performed at both $B=0$~T and $B=14$~T using a steady field applied parallel to the $c$-axis (normal to the NiO$_2$ plane) in a Teslatron cryostat (Oxford Instruments) and with a Synktek MCL lock-in amplifier.

\begin{figure}[h]
\centering
\includegraphics[width=1\textwidth]{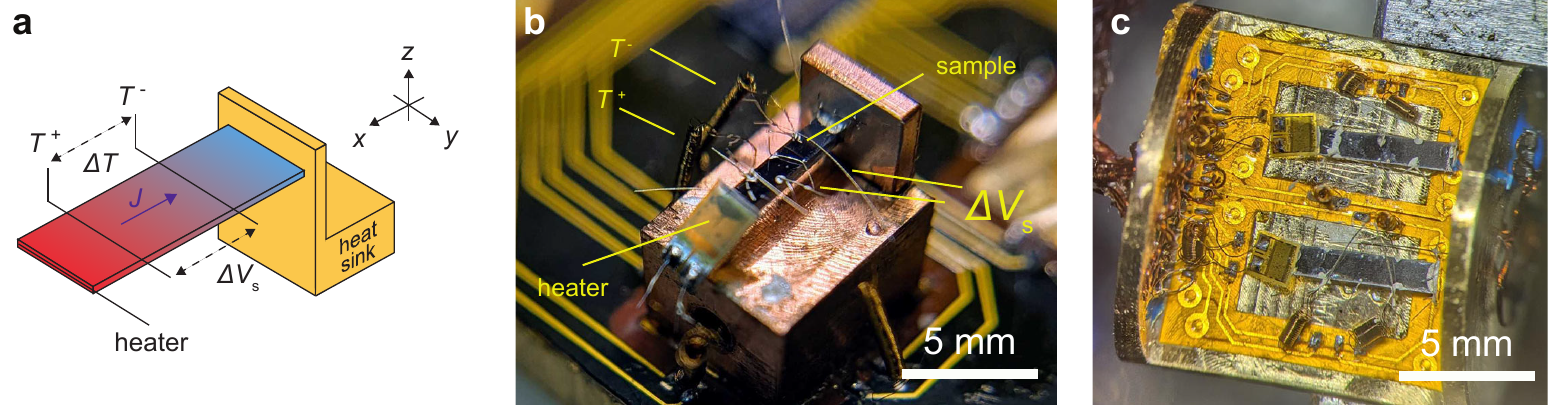}
\caption{
Seebeck experiment. $T^+$ is the temperature closest to the heater, $T^-$ is the temperature farthest from the heater, $\Delta T$ is the temperature gradient between the two, $\Delta V_{\rm S}$ is the voltage response of the sample and $J$ is the heat current. (\textbf{a}) Sketch; (\textbf{b}) École Polytechnique setup; (\textbf{c}) NHMFL setup.
}
\label{fig:seebeck_mount}
\end{figure}

\textbf{Thermoelectric measurements.}
To measure the Seebeck effect, we used a $2\omega$ AC method at low frequency below $2\omega = 1$~Hz, described in \cite{Gourgout2022Seebeck,grissonnanche2024}. For this, the heater (a 5~k$\Omega$ strain gauge) was glued directly on one end of the sample, the thermal gradient $\Delta T = T^+ - T^-$ measured with two type E thermocouples, and thermoelectric voltage $\Delta V_{\rm s}$ was probed on the longitudinal contacts used for resistivity in the same location as the thermocouples (Fig.~\ref{fig:seebeck_mount}). The sample was grounded on the opposite side to the heater.

Experiment was performed in a steady magnetic field $B$ applied parallel to the $c$-axis. At École Polytechnique, the Seebeck voltage was measured with a Synktek MCL lock-in amplifier while doing temperature steps. At the NHMFL, it was measured with a Stanford Research 860 lock-in amplifier and a maximum magnetic field of $41.5$ tesla while sweep up in temperature.

The Seebeck coefficient $S$ was obtained from :

\begin{equation}
S =  \frac{|\Delta V_{\rm s}| e^{i\phi_S}}{|\Delta T| e^{i\phi_T}}
\end{equation}

with $|\Delta V_{\rm s}|$ ($|\Delta T|$) the modulus of the Seebeck voltage (thermal gradient) measured by the lock-in, and $\phi_S$ ($\phi_T$) its phase, and $\phi_S \equiv \phi_T \pmod{\pi}$.

\begin{figure}[h]
\centering
\includegraphics[width=0.8\textwidth]{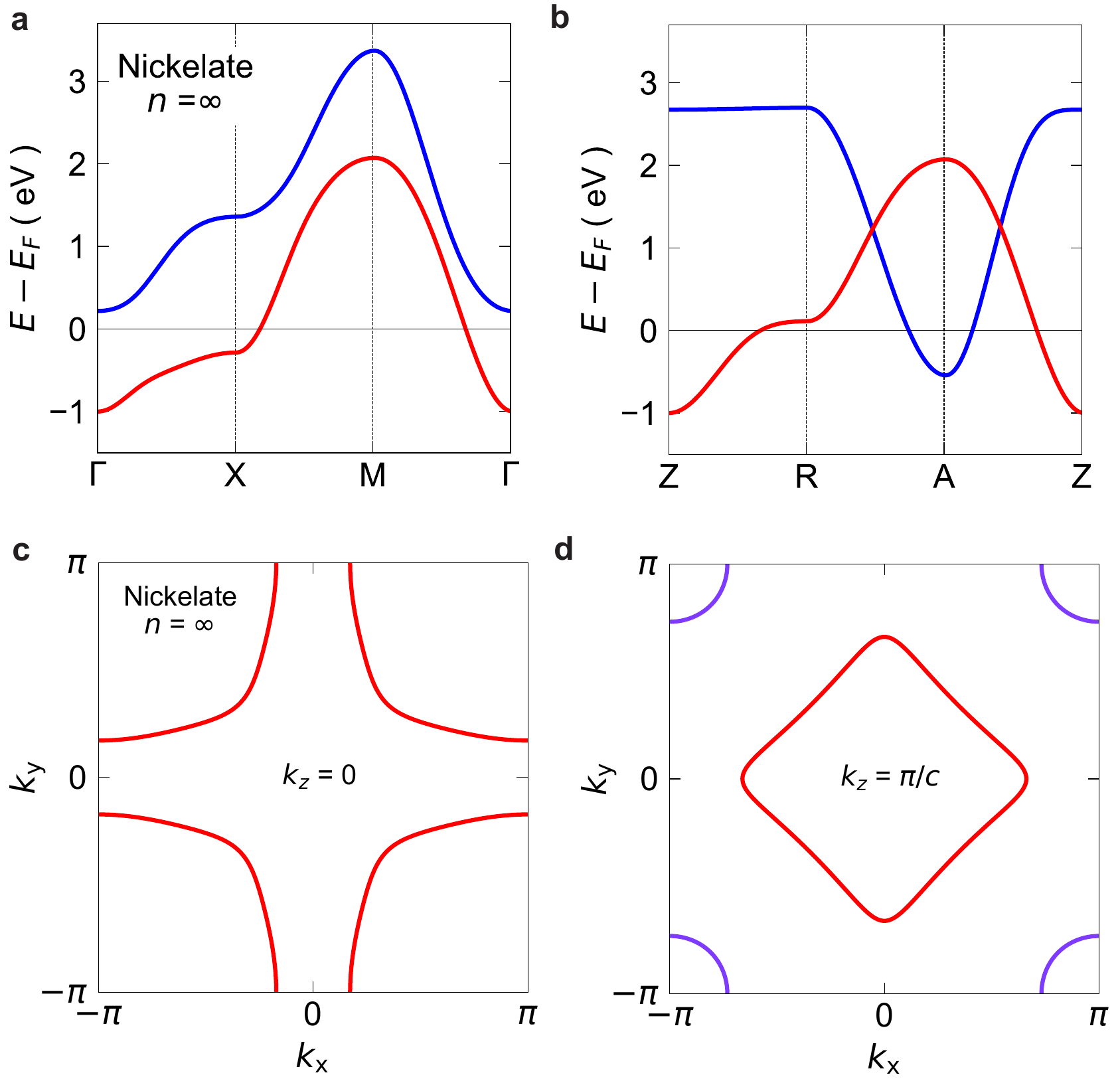}
\caption{
(\textbf{a,b}) Electronic dispersion of infinite nickelates based on the tight-binding model derived from ARPES data~\cite{Sun2025Electronic}, for (a) $k_z = 0$ and (b) $k_z=\pi/c$; (\textbf{c,d}) Corresponding Fermi surfaces for (c) $k_z = 0$ and (d) $k_z=\pi/c$.
}
\label{fig:band_structure_nickelates}
\end{figure}

\textbf{Seebeck coefficient calculations.} The Seebeck coefficient $S$ was calculated using Boltzmann transport theory in the relaxation time approximation following the same method and algorithm described in the cuprates~\cite{Gourgout2022Seebeck} and in $5$- and $3$-layer nickelates~\cite{grissonnanche2024}. In the high temperature regime, we use the constant scattering time approximation, which assumes that the energy dependence of the scattering rate is subdominant, as demonstrated for other nickelates~\cite{grissonnanche2024}. In this approximation, $S$ only depends on the band structure close to the Fermi level, and the scattering time $\tau$ drops out of the calculation. The band structure was obtained from a tight-binding model derived from ARPES data (Fig.~\ref{fig:band_structure_nickelates}) measured on LSNO $x=x^\star$ by Sun~\textit{et al.}~\cite{Sun2025Electronic}.

\textbf{Carrier density calculations.} To extract the carrier density $n_{\rm d}$ of the Ni-$d_{x^2-y^2}$ pocket as a function of doping $x$, we combine the Hall coefficient $R_{\rm H}$ in the $T\rightarrow 0$ limit with a two-band model involving $n_{\rm d}$ and $n_{\rm s}$, the carrier density associated with the Nd-$s$ pocket. The key assumption used to invert $R_{\rm H}$ and obtain $n_{\rm d}(x)$ is that $n_{\rm s}$ is approximately independent of doping. This is justified, to first order, by the three-dimensional and weakly correlated character of the Nd-$s$ pocket: (i) a two-band description captures the temperature dependence of the Hall effect in overdoped samples~\cite{Osada2025Systematic}, and (ii) DFT calculations indicate only a weak doping evolution of the $s$ pocket, insufficient on its own to account for the large change in $R_{\rm H}$ across $x^\star$~\cite{liu_2021}.

The Hall coefficient of a two-band model can be written as

\begin{equation}\label{eq:RH_full}
R_{\rm H} =  \frac{V}{e}\frac{n_{\rm d} \mu_d^2 - n_{\rm s} \mu_s^2 }{(n_{\rm d }\mu_d+n_{\rm s} \mu_s)^2}
\end{equation}

with  $V=3.962 \rm{\AA} \times 3.962 \rm{\AA} \times 3.268 \rm{\AA}$ the unit cell volume, $\mu_{\rm d}$ and $\mu_{\rm s}$ the mobility of each charge carrier, and $e$ the elementary charge.

By definition, the mobility $\mu = \frac{e\tau}{m^\star}$ with $\tau$ the scattering time and $m^\star$ the effective mass. In the $T\rightarrow 0$ limit, we assume that the mean free path $l_0$ at $T=0$ is limited by the averaged impurities distance, which translates as $l_0= v^d_F\tau_d = v^s_F\tau_s$. In this limit, the Hall coefficient simplifies as

\begin{equation}\label{eq:RH_reduced}
R_{\rm H}(T\rightarrow 0) =  \frac{V}{e}\frac{n_{\rm d} - n_{\rm s} r^2 }{(n_{\rm d }+n_{\rm s} r)^2}
\end{equation}

with $r= \frac{m_d v^d_F}{m_s v^s_F}$, and $m_d$ ($m_s$) the effective mass and $v^d_F$ ($v^s_F$) the averaged Fermi velocity on the Ni-$d_{x^2-y^2}$ pocket (Nd-$s$ pocket).

We obtain the effective masses and Fermi velocities directly from a tight-binding model fitted from the band structure measured by APRES~\cite{Sun2025Electronic} at $x=0.20$. We obtain a ratio $r = 3.3$. From inverting equation~\ref{eq:RH_reduced}, we obtain $n_{\rm d}$ as a function of doping $x$ (Fig.~\ref{fig:carrier_density_vs_doping}a).

\begin{figure}[h]
\centering
\includegraphics[width=1\textwidth]{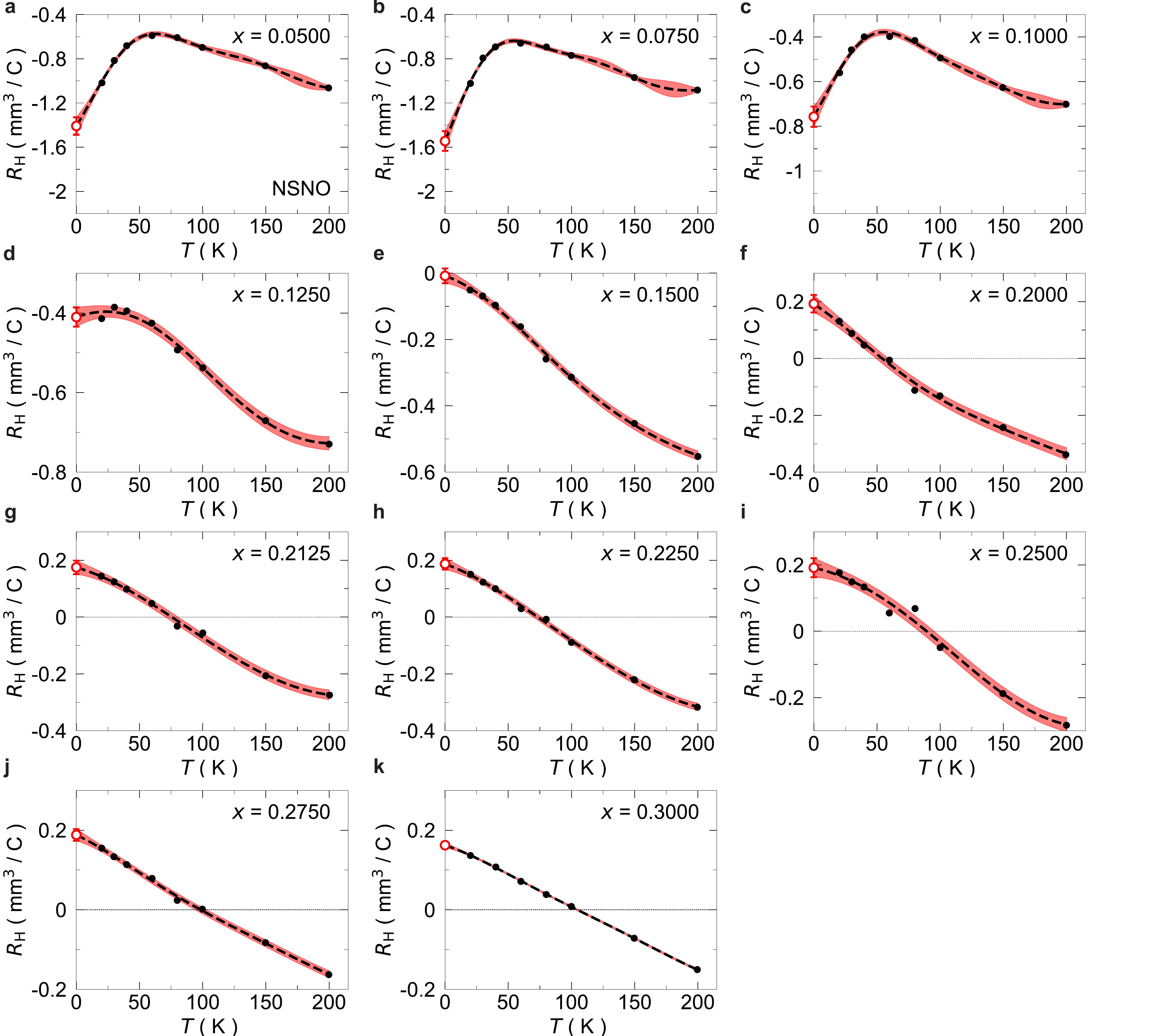}
\caption{
Hall effect $R_{\rm H}$ vs $T$ of Nd$_{1-x}$Sr$_x$NiO$_2$ (NSNO) at different doping $x$. Full points are reproduced from Lee~\textit{et al.}~\cite{lee_linear_character_2023}. Empty points are extrapolated at $T =0$~K through a Gaussian process regression. The red shading represents one standard deviation from the regression's mean prediction (dashed line).
}
\label{fig:hall_effect_vs_T}
\end{figure}

To extrapolate $R_{\rm H}$ in the $T \rightarrow 0$ limit from Ref.~\cite{lee_linear_character_2023}, we perform a Gaussian-process regression (GPR) using a radial-basis-function (RBF) kernel (Fig.~\ref{fig:hall_effect_vs_T}). For each doping, the kernel length scale and the noise level are optimized by maximizing the log marginal likelihood. The RBF length scale is constrained to the range $50$--$100$~K in order to preserve the influence of the lowest-temperature data points on the extrapolation. We include an additional white-noise kernel, with its variance constrained to $0.001$--$0.1$~mm$^3$/C, where the upper bound prevents the noise term from washing out the structure in $R_{\rm H}(T)$. Uncertainties are taken as one standard deviation of the GPR posterior.

In Fig.~\ref{fig:carrier_density_vs_doping}a, the error bar on $n_{\rm d}$ is defined as the difference between the value extracted from the posterior mean of $R_{\rm H}(T\rightarrow 0)$ and the value extracted from the corresponding lower one-standard-deviation prediction. Implementation details follow the \texttt{scikit-learn} GPR package documentation.

% % %-----------------------------------------%
% % %############# Acknowledgments ###########%
% % %-----------------------------------------%

\section*{Funding Statement}
Gaël Grissonnanche acknowledges support from the STeP2 no ANR-22-EXES-0013, QuantExt no ANR-23-CE30-0001-01, the École Polytechnique foundation, FrontQuant no ANR-24-CE97-0004 at Audace CEA No. ANR-24-RRII-0004. A portion of this work was performed at the National High Magnetic Field Laboratory, which is supported by National Science Foundation Cooperative Agreement No. DMR-2128556* and the State of Florida. Charles Iorio-Duval acknowledges support from the National Science and Engineering Council of Canada under a CGRS-M scholarship, the Fonds de Recherche du Québec under grant B1X-351563 and Mitacs Inc. under a Global Research Award.
% Benoît Fauqué, Kamran Behnia, Chung-Hou Chung

\section*{Author Contributions}
C.I.D., A.G., G.G. designed the study. W.S. and Y.F.N. grew the sample. F.P., J.L.S.G. and A.G. performed the proof of concept and preliminary experiments. C.I.D. and A.G. did the sample preparation and characterization. C.I.D., D.G. and G.G. performed the Seebeck experiment at the National High Magnetic Field Laboratory in Tallahassee. C.I.D, S.U.Z. performed the data analysis. C.I.D. and E.B.B performed the simulations. C.I.D and G.G. wrote the manuscript with input from all other co-authors. A.G. and G.G. supervised the project.

\section*{Competing Interests}
The authors declare no competing interests.

\section*{Additional information}
Correspondence and requests for materials should be addressed to G.G.

\section*{Extended Data}

%#################### Fig sup Comparison sample ####################%
%##################################################%

%#################### Figure ED - Seebeck mount ####################%
%##################################################%

%#################### Fig ED - Band structure nickelates ####################%
%##################################################%

%#################### Figure ED - Hall vs T ####################%
%##################################################%

%#################### Fig ED - RH(0) vs x ####################%
\begin{figure}[h]
\centering
\includegraphics[width=0.6\textwidth]{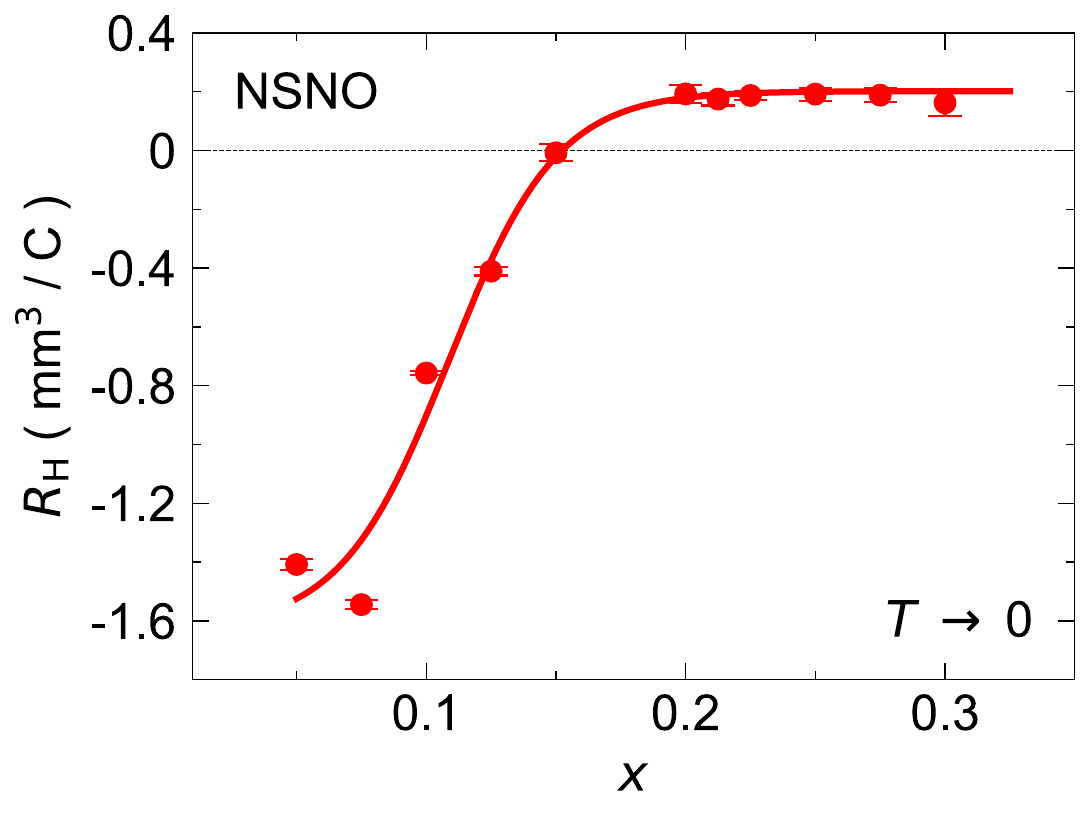}
\caption{
Hall effect $R_{\rm H}$ vs doping $x$ of Nd$_{1-x}$Sr$_x$NiO$_2$ (NSNO) in the limit $T\rightarrow 0$. Data is extrapolated from Lee~\textit{et al.}~\cite{lee_linear_character_2023} (Fig.~\ref{fig:hall_effect_vs_T}). Lines are a guide to the eye.
}
\label{fig:hall_effect_0_vs_x}
\end{figure}
%##################################################%

% %-----------------------------------------%
% %############### Biblio ##################%
% %-----------------------------------------%

% % Create the reference section using BibTeX:

\end{document}